\documentclass[prl,preprint,amsmath,showpacs,groupedaddress,superscriptaddress]{revtex4}
\usepackage[english]{babel}
\usepackage[applemac]{inputenc}
\usepackage{latexsym}
\usepackage{hyperref}
\usepackage{graphicx}
\usepackage{epstopdf}
\usepackage{times} 

\begin{document}




\title {Magneto-optical evidence of double exchange in a percolating lattice}

\author{G. Caimi}
\affiliation{Laboratorium f\"ur Festk\"orperphysik, ETH Z\"urich,
CH-8093 Z\"urich, Switzerland}

\author{A. Perucchi}
\affiliation{Laboratorium f\"ur Festk\"orperphysik, ETH Z\"urich,
CH-8093 Z\"urich, Switzerland}

\author{L. Degiorgi}
\affiliation{Laboratorium f\"ur Festk\"orperphysik, ETH Z\"urich,
CH-8093 Z\"urich, Switzerland}

\author{H.~R. Ott}
\affiliation{Laboratorium f\"ur Festk\"orperphysik, ETH Z\"urich,
CH-8093 Z\"urich, Switzerland}

\author{V.~M. Pereira}
\affiliation{Departement of Physics, Boston University, 590 Commonwealth Av., Boston MA 02215, U.S.A.}
\affiliation{CFP and Departamento de F{\'\i}sica, Faculdade de Ci\^encias Universidade de Porto, 4169-007 Porto, Portugal}

\author{A.~H. Castro Neto}
\affiliation{Departement of Physics, Boston University, 590 Commonwealth Av., Boston MA 02215, U.S.A.}

\author{A.~D. Bianchi}
\affiliation{National High Magnetic Field Laboratory, Florida State University, Tallahassee FL 32306, U.S.A.}
\affiliation{Hochfeldlabor Dresden, Forschungszentrum Rossendort, Postfach 510119, D-01314 Dresden, Germany}

\author{Z. Fisk}
\affiliation{Departement of Physics, University of California, Davis CA 95616, U.S.A.}\

\date{\today}

\begin{abstract}
Substituting $Eu$ by $Ca$ in ferromagnetic $EuB_6$ leads to a percolation limited magnetic ordering. We present and discuss magneto-optical data of the $Eu_{1-x}Ca_{x}B_6$ series, based on measurements of the reflectivity $R(\omega)$ from the far infrared up to the ultraviolet, as a function of temperature and magnetic field. Via the Kramers-Kronig transformation of $R(\omega)$ we extract the complete absorption spectra of samples with different values of $x$. The change of the spectral weight in the Drude component by increasing the magnetic field agrees with a scenario based on the double exchange model, and suggests a crossover from a ferromagnetic metal to a ferromagnetic Anderson insulator upon increasing $Ca$-content at low temperatures.
\end{abstract}
\maketitle

Because of the potential technological applications, materials exhibiting colossal magnetoresistive (CMR) effects are of high current interest in solid state physics. Europium hexaboride ($EuB_{6}$) and the well known manganites, for which the onset of ferromagnetism is accompanied by a dramatic reduction of the electrical resistivity, are primary examples, that have intensively been studied. Although a lot of experimental data on the magnetic and electronic properties is available, a complete understanding of the underlying physical mechanisms is still lacking.

We concentrate on the series of cubic $Eu_{1-x}Ca_{x}B_{6}$, which displays interesting correlations between magnetic, transport and optical properties. The high temperature electronic transport of $EuB_6$, a ferromagnet (FM) below $T_C\sim 12$~K, relies on a small effective electron density \cite{degiorgi97}. The magnetic properties are dominated by the half-filled $4f$ shell of divalent $Eu$, which accounts for the measured magnetic moment of $7\mu_{B}$ per formula-unit \cite{henggeler}. The strong coupling of transport properties to the magnetization was revealed by measurements of the magneto-optical properties, which unveiled a substantial blue shift of the plasma edge in the optical reflectivity with decreasing temperature and increasing magnetic field \cite{broderick02}. The remarkable results correlate with the precipitous drop in the dc resistivity just below $T_C$, and the large negative magnetoresistance observed near $T_C$ \cite{sullow,Paschen:2000,wigger05}. 

A variety of models was invoked to address those properties. Full potential spin-polarized self-consistent electronic band structure calculations suggest a semi-metallic or degenerate semiconductor state of $EuB_6$ \cite{ghosh}. This view does not agree with results  from Angular-Resolved Photoemission Spectroscopy (ARPES) that reveal a considerable gap between valence and conduction bands \cite{denlinger}. Our magneto-optical Kerr effect data set \cite{broderickMOKE02} implies that the inclusion of a Drude contribution is essential in interpreting the experimental optical data for $EuB_6$. Lin and Millis were able to reproduce $T_C$, the enhancement of the plasma frequency in the ordered phase \cite{degiorgi97,broderick02} and the specific heat \cite{degiorgi97} with a calculation based on the dynamical mean-field approximation \cite{lin}. The major finding of this latter approach establishes that the splitting of the $Eu$ derived conduction band, due to the coupling to core electrons at each $Eu$ site, causes the phase transition. Another model, based on the formation of magnetic polarons \cite{calderon}, suggests that a Hund-type coupling of holes to local spins and magnetic ordering via the Ruderman-Kittel-Kasuya-Yoshida (RKKY) interaction are responsible for the ferromagnetic phase transition and CMR. Wigger \textit{et al.} \cite{wigger05}, on the other hand, reproduced the negative magnetoresistance in the paramagnetic regime by a model in which the spin disorder scattering of the itinerant electrons is reduced by the applied magnetic field, i.e., by the increasing magnetization, and also accounts for the results of de Haas-van Alphen, optical reflectivity and ARPES measurements. 

The intimate relation between magnetization and electronic conductivity also emerged from experiments on the $Eu_{1-x}Ca_{x}B_{6}$ series \cite{wigger02,perucchi04,wigger04,wigger051}. The $Ca$-substitution leads to significant changes of the magnetic and electronic properties. The FM transition temperature decreases with increasing $Ca$ content and stoichiometric $CaB_6$ exhibits no magnetic order \cite{wigger04,wigger051}. Evidence was also established for a spin-filter effect in the charge transport and dynamics for $x=0.4$ \cite{wigger02,perucchi04}. The electrical transport properties seem to be governed by percolation-type phenomena across the $Eu_{1-x}Ca_{x}B_{6}$ series \cite{wigger04,wigger051}. At about $x=0.7$, which coincides with the site-percolation limit in a simple cubic lattice, the long range order disappears and for $0.7<x<0.9$ a spin-glass type ground state is adopted \cite{wigger051}. 

A recently published approach to explain the behavior of $EuB_6$, as well as offering specific predictions for the electronic properties of the $Eu_{1-x}Ca_{x}B_{6}$ series, is based on a double-exchange scenario \cite{pereira}. This scenario may be regarded as an effective theory for the Kondo lattice problem in the limit of a very small number of carriers. The reduced itinerant carrier concentration places the Fermi level near a magnetization dependent mobility edge, which emerges in the spectral density because of the disordered spin background and/or $Ca$-doping. An FM metal to insulator crossover is expected as a function of the position of the Fermi level with respect to the mobility edge, which can be tuned by the $Ca$-content \cite{pereira}. The model also addresses \cite{pereira05} the region of stability of magnetic polarons in the paramagnetic phase near $T_C$ \cite{nyhus97}.

The goal of this paper is to present and analyze our magneto-optical data of the $Eu_{1-x}Ca_{x}B_{6}$ series. Replacing $Eu$ by $Ca$ has direct consequences on the electrodynamic response. It influences the distribution of the spectral weight between the metallic (Drude) component and excitations at non-zero energy in the absorption spectrum for different $Ca$-contents at different temperatures and magnetic fields. Our analysis provides support for the phase diagram that emerges from the double-exchange model predictions \cite{pereira}. The boundary between the metallic and the insulating ferromagnetic state at low temperatures is found to be close to $x=0.5$.

$Eu_{1-x}Ca_{x}B_{6}$ single crystals of high structural quality were prepared by solution growth from $Al$ flux, using the necessary high-purity elements as starting materials \cite{wigger04,wigger051}. The optical reflectivity $R(\omega)$ was measured in a broad spectral range from the far infrared (FIR) to the ultraviolet, and as a function of both temperature (1.6-300 K) and magnetic field (0-7 T) \cite{degiorgi97,broderick02}. Because of the broad coverage of spectral range, $R(\omega)$ spectra allowed for a reliable Kramers-Kronig (KK) transformation providing the complete absorption spectrum represented by the real part $\sigma_1(\omega)$ of the optical conductivity \cite{wooten,dressel}. At low frequencies, i.e., below our low frequency experimental limit of about 30 $\mathrm{cm}^{-1}$, $R(\omega)$ was extended using the Hagen-Rubens extrapolation, and inserting the $\sigma_{dc}$ values obtained with d.c. transport measurements \cite{wigger04,wigger05,wigger02}. Above the highest measurable frequency, $R(\omega)$ was extended into the electronic continuum with the standard extrapolations $R(\omega)\sim \omega^{-s}$ ($2<s<4$) \cite{wooten,dressel}.

\begin{figure}[t]
  \begin{center}
		\includegraphics[width=0.9\columnwidth]{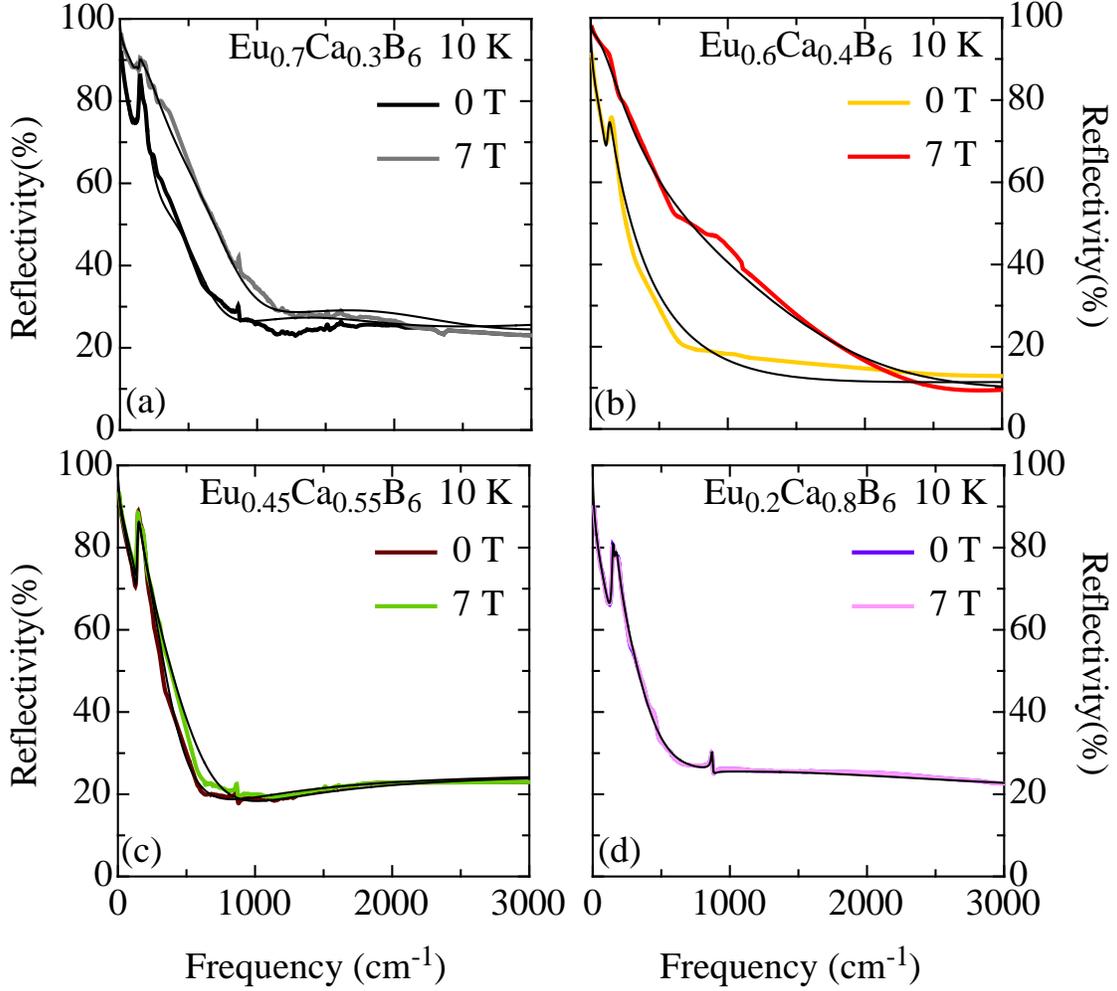}
		\caption{(color online) Reflectivity spectra $R(\omega)$ of $Eu_{1-x}Ca_{x}B_{6}$  in the infrared spectral range at $T= 10$~K $(T>T_C)$, for magnetic fields of 0~T and 7~T. Note that for $x=0.8$ the 0 and 7 T spectra are identical. The thin lines represent the Lorentz-Drude fit (see text).}
		\label{fig1}
	\end{center}
\end{figure}

The temperature and magnetic field dependence of $R(\omega)$ for $EuB_6$ was already presented and discussed in Refs. \onlinecite{degiorgi97} and \onlinecite{broderick02}. Here, we complement those results with data for $x=$0.3, 0.4 (Ref. \onlinecite{perucchi04}), 0.55 and 0.8. For the presentation in Fig.~\ref{fig1}, we chose spectra that were recorded at $10\,$~K (i.e., $T>T_C$) for all compounds, and in magnetic fields of 0~T and 7~T. $R(\omega)$ is progressively enhanced with increasing magnetic field for $x=0.3$ and $0.4$, while for $x=0.55$ and $0.8$, basically no field dependence was registered. Although $R(\omega)$ exhibits metallic character for all these $Ca$-contents, the onset of the plasma edge in $R(\omega)$ is quite broad in all compounds. This is distinctly different from the previously observed rapid increase of $R(\omega)$ with decreasing $\omega$ for $EuB_6$ (Refs. \onlinecite{degiorgi97} and \onlinecite{broderick02}).

\begin{figure}[t]
	\begin{center}
  	\includegraphics[width=0.7\columnwidth]{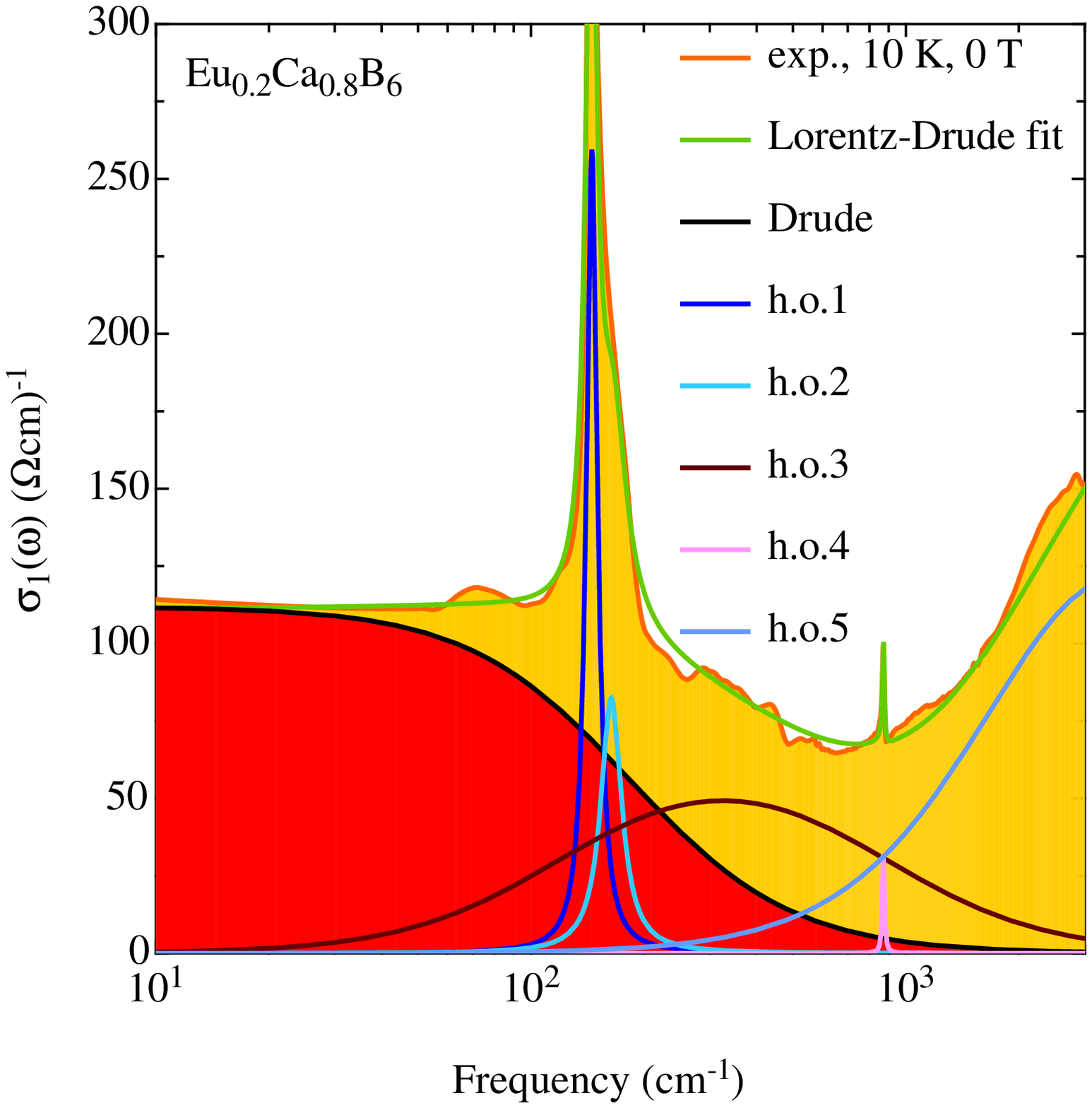}
		\caption{(color online) (a) Real part, $\sigma_1(\omega)$, of the optical conductivity of $Eu_{0.2}Ca_{0.8}B_{6}$ at 10~K and 0~T. The total Lorentz-Drude fit as well as the fit components are displayed, and one can appreciate the very satisfying fit quality. The colored red area identifies the Drude spectral weight, while the sum of the red and yellow areas represents the total spectral weight encountered in $\sigma_1(\omega)$.}
		\label{fig2}
	\end{center}
\end{figure}

In order to analyze the spectra, we apply the phenomenological Lorentz-Drude approach \cite{wooten,dressel}. The fit procedure consists in considering a Drude term for the zero-frequency mode excitation due to the itinerant charge carriers, and an appropriate number of Lorentz harmonic oscillators (h.o.) for the absorptions at non-zero energies. The latter components represent phonon modes, electronic interband transitions and localized-state excitations.  Although of phenomenological character, this approach allows for the evaluation of the spectral-weight distribution among the various components of the absorption spectrum. For all $Ca$-substitutions, we consistently reproduce each spectrum for any combination of temperature and magnetic field with the same fit procedure and the same number of fit components. As an example, we show $\sigma_1(\omega)$ at 10~K and 0~T for $x=0.8$ in Fig.~\ref{fig2}. The individual components to the fit may readily be identified. Both Figs.~\ref{fig1} and \ref{fig2} demonstrate that in this way the raw experimental data for $R(\omega)$ and the resulting $\sigma_1(\omega)$ can be reproduced very well. We can thus disentangle the Drude spectral weight corresponding to the squared plasma frequency from the spectral weight associated with the excitations at non zero energies. The total spectral weight encountered in the excitations at higher energies is proportional to the sum of the squared mode strengths \cite{wooten,dressel}.

\begin{figure}[t]
	\begin{center}
		\includegraphics[width=0.7\columnwidth]{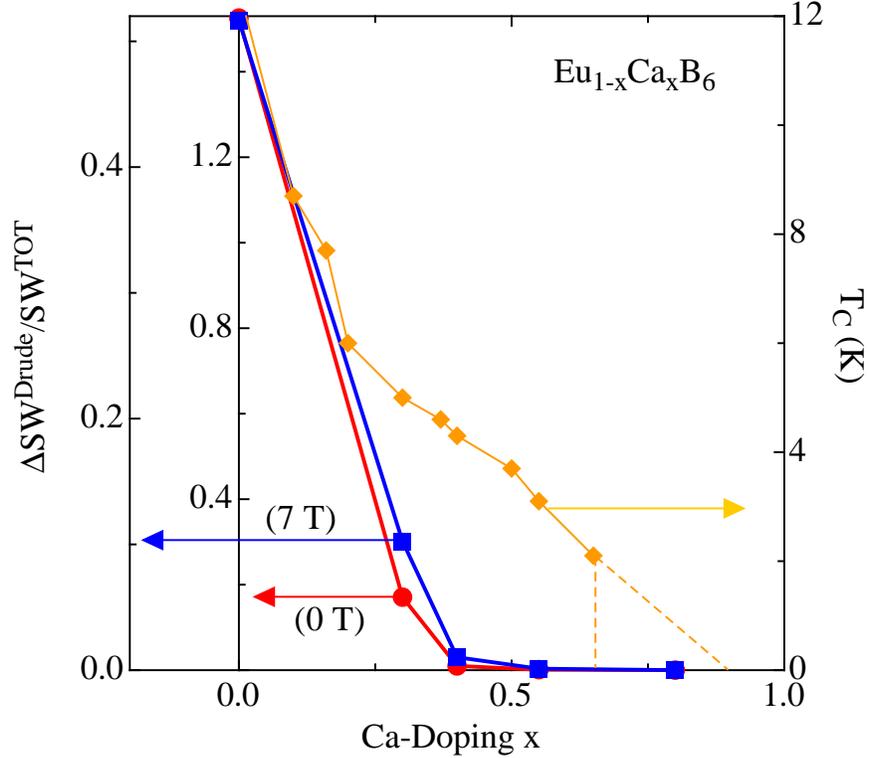}
		\caption{(color online) Dependence of the Curie temperature, $T_C$, on $Ca$-doping \cite{wigger04,wigger051}, compared to $\Delta SW^{Drude}/SW^{TOT}$, the change of the spectral weight in the Drude component, normalized by the total spectral weight (at 0 or 7~T) encountered in $\sigma_1(\omega)$ up to $\omega_c$. The dashed lines for $T_C(x)$ above $x\sim 0.7$ schematically define the interval of $T$ and $Ca$-content, where a cluster spin glass phase was established \cite{wigger04,wigger051}.}
		\label{fig3}
	\end{center}
\end{figure}

Because of the specific model prediction \cite{pereira} of a FM metal to insulator crossover, our primary aim is to evaluate the change of spectral weight ($SW$) of the metallic component of $\sigma_1(\omega)$ (red area in Fig. 2) at 10 K between 0 and 7 T. This difference is calculated on the basis of our Lorentz-Drude results and is defined as \cite{note}: $\Delta SW^{Drude}=SW^{Drude}$(10 K, 7 T)- $SW^{Drude}$(10 K, 0 T). The magnetic field of 7 T is high enough to drive the system into magnetic saturation for all values of $x$, such that  $\sigma_1(\omega)$ at 7~T reflects the maximum metallicity reached by increasing the magnetic field. The temperature of 10 K was chosen to be close enough to, but above, $T_C$ for each compound \cite{comment}. In order to allow for a comparison for different $Ca$-contents, we renormalize the change of the Drude spectral weight $\Delta SW^{Drude}$ by the total spectral weight (red plus yellow areas in Fig.~\ref{fig2}) encountered in $\sigma_1(\omega)$ up to about $\omega_c=1$~eV, at either 0 or 7~T, i.e. $SW^{TOT}$(10 K, 0 or 7 T) (Ref. \onlinecite{note}). The cut-off frequency $\omega_c$ coincides approximately with the onset of the high-frequency electronic interband transitions, where both the temperature and the magnetic field dependence in $\sigma_1(\omega)$ vanish. Figure ~\ref{fig3} displays the variation of the normalized Drude spectral weight (i.e., $\Delta SW^{Drude}/SW^{TOT}$) as a function of $x$, in comparison with the variation of $T_C$ for the $Eu_{1-x}Ca_{x}B_6$ series  \cite{wigger04,wigger051}. The ranges on the vertical axes were chosen such that $T_C(x=0)$ coincides with the renormalized changes of the Drude spectral weight, inserting $SW^{TOT}$ for either 0 or 7 T. For both values of $SW^{TOT}$, we obtain the same type of variation with $x$. $\Delta SW^{Drude}$ decreases sharply between $x=0$ and 0.3, reaching zero at approximately $50\%$ $Ca$-content. 

The $x$-dependence of $\Delta SW^{Drude}$ reveals the reduction of the maximum of itinerant charge carriers with increasing $x$ in $Eu_{1-x}Ca_{x}B_{6}$, previously indicated by results of resistivity \cite{Paschen:2000,wigger02}, Hall effect \cite{wigger04} and optical response \cite{degiorgi97,broderick02,perucchi04} measurements. The field-induced enhancement of $\Delta SW^{Drude}$ as a function of $x$ and its correlation with the evolution of $T_C(x)$ is fully consistent with the microscopic mechanisms that are considered in the low-density double-exchange model applied to a percolating lattice \cite{pereira}. The crucial detail of this description of the interaction between the conduction electrons and the \textit{classical} ($S=7/2$) localized magnetic moments is not the usual strong Hund coupling condition ($J_H\gg t$), but rather the extremely reduced density of itinerant carriers that permits the high-energy electronic states to be projected out. Part of the experimental evidence \cite{sullow,Paschen:2000,wigger05,wigger02,perucchi04,wigger04,wigger051} for the strong interplay between transport and magnetisation in the hexaborides is then the consequence of Anderson localization effects. In this sense, the magneto-transport in $Eu_{1-x}Ca_{x}B_{6}$ is dominated by a magnetization dependent source of disorder in the random magnetic background in which the electrons itinerate. This is quantified by the position of the mobility edge, $E_C(\mathcal M)$, relative to the Fermi energy, $E_F(\mathcal M)$ \cite{foot}. The reduced electronic densities, characteristic of the $Eu_{1-x}Ca_{x}B_{6}$ series, place these two energies very close to each other. Direct experimental consequences of this are the marked enhancement of the itinerant carrier density upon driving the system into a ferromagnetic regime by either reducing $T$ to below $T_C$ or by applying an external magnetic field. With $E_C(\mathcal M)$ shifting towards the band edge upon increasing $\mathcal M$, more extended (metallic) states are occupied in the system.

The microscopic model of Ref. \onlinecite{pereira} considers an optical sum rule that includes only the Drude contribution due to the extended states:
$%
\omega_p^2(\mathcal{M}) \propto \int_{E_C(\mathcal M)}^\infty N(E,\mathcal{M}) f(E)EdE \,,
$
with $N(E,\mathcal{M})$ as the magnetization dependent electronic density of states, and $f(E)$ the Fermi-Dirac function. Upon replacing $Eu$ by $Ca$, the $Eu$ sublattice and consequently the magnetic subsystem are diluted. Given the isovalency of $Ca$ and $Eu$, from the electronic point of view, the main effect of this is a strong enhancement of the electronic disorder, and thus $E_C(\mathcal M,x)$ will also depend on the $Ca$ content. For this reason, with increasing $x$ less extended states are occupied, and once $E_C(\mathcal M,x_{MI}) = E_F(\mathcal M,x_{MI})$, a metal-insulator transition (at $T=0$) or a metal-insulator crossover ($T>0$) emerges. Figure~\ref{fig3} shows that the spectral weight of the Drude term in $\sigma_1(\omega)$ is enhanced when the system is driven from a paramagnetic metallic state into full polarization up to $x \lesssim 0.4$. The progressive reduction of this enhancement with $x$ reflects the drift of $E_C$ and consequently less metallic conduction. For higher $Ca$ concentrations, the Drude spectral weight is insensitive to the spin polarization, a signal that the mobility edge $E_C(\mathcal M=1)$ went past the Fermi energy and thus the polarization no longer \textit{releases} any of the localized states. The tail-like behavior of $\Delta SW^{Drude}/SW^{TOT}$ in Fig. 3 for $x \gtrsim 0.4 $ can be understood as originating from the non-zero temperature excitations of carriers across the mobility gap, $E_C - E_F$ (all measurements were made at $T>T_C$). Both this interpretation and the features of the optical response are perfectly consistent with the variation of the number of carriers in the paramagnetic phase, $n_0(x)$, presented in Ref.~\onlinecite{wigger04}. It is also significant that the results do not depend on the magnetic field used for the normalization of $\Delta SW^{Drude}$, as evidenced by the two curves in Fig.~\ref{fig3}, where $\Delta SW^{Drude}$ is normalized by $SW^{TOT}$ at either 0 or 7~T. 

With respect to magnetic order, Fig. 3 shows that the Curie temperature decreases with $x$, as expected if site percolation is important. Unlike the Drude spectral weight, long-range magnetic order survives until the $Ca$ concentration coincides with the threshold of the simple cubic magnetic lattice site percolation. This is in agreement with the double-exchange scenario in which the effective magnetic coupling is the result of the electron itinerancy among sites with localized moments. This magnetic coupling, local in nature, is present even in the localized regime as long as typical localization lengths allow wavefunctions to spread over nearest neighbors. From the optical point of view, Fig.~\ref{fig3} perfectly reflects the phase diagram predicted in Ref.~\onlinecite{pereira}. Up to $x_{MI}\simeq 0.4$ one has a metallic ferromagnet; for higher $Ca$ concentration the system remains a ferromagnetic Anderson insulator until it reaches the percolation threshold. Near and above the percolation threshold the number of disconnected $Eu$-rich magnetic clusters becomes significant. Even though the tendency should be towards ferromagnetism, it is not surprising that the regime above $x\gtrsim0.7$ (beyond percolation, and at very low temperatures) seems to be characterized by glassy magnetism \cite{wigger051}, on account of the possible presence of superparamagnetic clusters and competing dipolar interactions at such extreme dilutions of the magnetic moments \cite{comment2}.

Finally the $Ca$-driven metal-insulator transition found in transport measurements, now corroborated with our magneto-optical measurements, is a detail of paramount relevance in the phase diagram of $Eu_{1-x}Ca_xB_6$. Disorder is one of the essential building blocks of the double-exchange interpretation and the current confirmation of the phase diagram foreseen in Ref.~\onlinecite{pereira} certainly underlines the significance of localization effects in the description of these compounds. 

In conclusion, our magneto-optical investigations on the $Eu_{1-x}Ca_{x}B_6$ series reveal a phase diagram in support of a scenario based on the close proximity of the Fermi level and a magnetization dependent mobility edge. A ferromagnetic metal-insulator crossover occurs upon increasing the $Ca$-content in $Eu_{1-x}Ca_xB_6$ to above $x_{MI}$. The magneto-optical data suggest a critical $Ca$-content \mbox{$x_{MI}\sim 0.5$}. 

\acknowledgments
The authors wish to thank J. M\"uller for technical help, and D. Basov, R. Monnier and S. Broderick for fruitful discussions.
V.~M.~P. acknowledges the support of FCT, through grant SFRH/BD/4655/2001, and thanks Boston University for the hospitality.
A.~H.~C.~N. was supported through NSF grant DMR-0343790.
This work has been supported by the Swiss National Foundation for the Scientific Research, within the NCCR research pool MaNEP. Work at UC Davies benefited from financial support of the US National Science Foundation under contract DMR-0433560.

\bibliographystyle{prsty}

\begin{thebibliography}{10}
\bibitem{degiorgi97}
L. Degiorgi \textit{et al.}, Phys. Rev. Lett. {\bf79}, 5134 (1997).
\bibitem{henggeler}
W. Henggeler \textit{et al.}, Solid State Commun. {\bf108}, 929 (1998).
\bibitem{broderick02}
S. Broderick \textit{et al.}, Phys. Rev.  B {\bf65}, 121102(R) (2002).
\bibitem{sullow}
S. S\"ullow \textit{et al.}, Phys. Rev.  B {\bf62}, 11626 (2000).
\bibitem{Paschen:2000} 
S. Paschen \textit{et al.}, Phys. Rev. B \textbf{61}, 4174 (2000).
\bibitem{wigger05}
G.A. Wigger \textit{et al.}, Phys. Rev.  B {\bf69}, 125118 (2004).
\bibitem{ghosh}
D.B. Ghosh \textit{et al.}, cond-mat/0406706.
\bibitem{denlinger}
J.D. Denlinger \textit{et al.}, Phys. Rev. Lett. {\bf 89}, 157601 (2002).
\bibitem{broderickMOKE02}
S. Broderick \textit{et al.}, Eur. Phys. J. B {\bf 27}, 3 (2002).
\bibitem{lin}
C. Lin and A.J. Millis, Phys. Rev. B {\bf 71}, 075111 (2005).
\bibitem{calderon}
M.J. Calderon \textit{et al.}, Phys. Rev. B {\bf 70}, 092408 (2004).
\bibitem{wigger02}
G.A. Wigger \textit{et al.}, Phys. Rev.  B {\bf66}, 212410 (2002).
\bibitem{perucchi04}
A. Perucchi \textit{et al.}, Phys. Rev. Lett. {\bf 92}, 067401 (2004).
\bibitem{wigger04}
G.A. Wigger \textit{et al.}, Phys. Rev. Lett. {\bf93}, 147203 (2004).
\bibitem{wigger051}
G.A. Wigger \textit{et al.}, Eur. Phys. J. B {\bf46}, 231 (2005).
\bibitem{pereira}
V.M. Pereira \textit{et al.} Phys. Rev. Lett. \textbf{93}, 147202 (2004).
\bibitem{pereira05}
V.M. Pereira \textit{et al.} cond-mat/0505741.
\bibitem{nyhus97}
P. Nyhus \textit{et al.} Phys. Rev. B \textbf{56}, 2717 (1997).
\bibitem{wooten}
F. Wooten, in {\itshape Optical Properties of Solids}, Academic Press, New York (1972).
\bibitem{dressel}
M. Dressel and G. Gr\"uner, in {\itshape Electrodynamics of Solids}, Cambridge University Press (2002).
\bibitem{note}
$SW^{Drude}\sim \omega_p^2$ and $SW^{TOT}\sim \omega_p^2 + \sum_{j} \omega_{pj}^2$, where $\omega_p$ is the plasma frequency of the Drude term and $\omega_{pj}$ are the mode strength of the Lorentz harmonic oscillators covering the spectral range up to $\omega_c$. An equivalent procedure for calculating $SW^{Drude}$ and $SW^{TOT}$ consists in integrating $\sigma_1(\omega)$ up to about 0.15~eV for the effective Drude spectral weight, and up to about $\omega_c$ for the total weight.
\bibitem{comment}
We confirmed that any choice of temperature $T_C<T<10$ K for each $Ca$-content leads to equivalent results.
\bibitem{foot}
The band curvature also depends on the strength of disorder.
\bibitem{comment2}
In Ref. \onlinecite{pereira}, the scattering is due to spin disorder and also from the disorder induced by the random position of atoms. The treatment of scattering processes is more involved in the case of the spin glass phase close to percolation. This requires a detailed understanding of the spin glass nature, an issue which is beyond the scope of the theory \cite{pereira}.
\end{thebibliography}

\end{document}